\documentclass[twocolumn,showpacs,amsmath,amssymb]{revtex4-1}
\usepackage{bm}
\usepackage[T1]{fontenc}
\usepackage{hyperref}
\input{epsf}

\usepackage{graphicx}
\usepackage{epstopdf}
\usepackage{array}
\usepackage{longtable}
\usepackage{rotating,booktabs}
\usepackage{booktabs,threeparttable}
\usepackage{bm}
\usepackage{float}
\usepackage{amsmath}
\usepackage{gensymb}
\usepackage{multirow}
\usepackage{epsfig}
\usepackage{color}

\begin{document}
 \title{Precision calculation of hyperfine-structure constants for extracting nuclear quadrupole moment of $^{43}$Ca}
\author{Yong-Bo Tang}
\affiliation {College of Engineering Physics, Shenzhen Technology University, Shenzhen, 518118, China}
\email{tangyongbo@sztu.edu.cn}
\date{\today}

\begin{abstract}
There have been several reported values for the nuclear quadrupole moment of $^{43}$Ca, but significant discrepancies exist among these reported values, ranging from \(-0.0408(8)\)~b to \(-0.065(20)\)~b. In this work, we performed an accurate calculation of the electric field gradients of the \(4s4p~^3\!P_{1}\), \(4s4p~^3\!P_{2}\) and \(4s3d~^1\!D_2\) states in the $^{43}$Ca atom using a hybrid method. This hybrid method integrates the advantages of the configuration interaction method and the coupled-cluster method, and can simultaneously account for core-core, core-valence, and valence-valence correlations.
By combining our calculated results with the  experimental
values of the electric quadrupole hyperfine-structure constants of these three states, an accurate and reliable
nuclear quadrupole moment of $^{43}$Ca was determined to be \(-0.0479(6)\)~b, which could be recommended as a reference for \(^{43}\text{Ca}\).
\end{abstract}

\maketitle
\section{Introduction}\label{sec1}
Atomic nuclei with a nuclear spin greater than $\frac{1}{2}$ possess an electric quadrupole moment $Q$. It is a fundamental parameter used to describe the degree of deviation of the nuclear charge distribution from spherical symmetry. This parameter plays a significant role in many research areas~\cite{Neyens2003RPP,Heyde2011RMP,Campbell2016PRNP,Pekka2018MP,Sinitsyn2012PRL}.
For instance, the nuclear quadrupole moment serves as a unique and excellent tool for studying nuclear deformation and shape coexistence, especially for exotic nuclei near the protons drip-line~\cite{Neyens2003RPP,Heyde2011RMP,Campbell2016PRNP}. In the study of molecular dynamics, an accurate understanding of the nuclear quadrupole moment is required for systems in which nuclear quadrupole effects determine the spin-lattice relaxation time~\cite{Pekka2018MP}. Moreover, the nuclear quadrupole moment can be employed as a microscopic probe to explore the motion of atomic tunneling systems in amorphous solids~\cite{Sinitsyn2012PRL}.

Although well-established benchmark values exist for the magnetic dipole moments of many nuclei, precise reference values for the electric quadrupole moments of numerous nuclei remain scarce~\cite{Pekka2018MP,Stone2005ADNDT}. Experimental measurements of the electric quadrupole hyperfine-structure (HFS) constant, in conjunction with calculated electric field gradient, can be used to determine nuclear quadrupole moment. This approach is independent of nuclear theory and stands out as one of the most precise methods for determining the nuclear electric quadrupole moment $Q$. The electric quadrupole moments $Q$ of some nuclei have been determined using this approach~\cite{Singh2012pra,Sahoo2013pra,Bi2018pra,Lu2019pra,Li2021pra,li2021ab,Porsev2021prl,Papoulia2021pra,Skripnikov2021prc,Skripnikov2024prc,Zhang2025pra}.
The aim of the present work is to apply this method to determine the nuclear quadrupole moments of $^{43}$Ca.

 Theoretically, the nuclear quadrupole moment of $^{43}$Ca can be derived through computations and measurements for neutral $^{43}$Ca atom or any $^{43}$Ca ion.
 To the best of our knowledge, the hyperfine-structures of the \(4s4p~^3\!P_{1}\), \(4s4p~^3\!P_{2}\) and \(4s3d~^1\!D_2\) states in Ca atom~\cite{Grundevik1979PRL, Arnold1981HI,Aydin1982ZPA}, as well as the $3d_{5/2}$ state in $^{43}$Ca$^{+}$~\cite{Benhelm2007PRA}, have been measured accurately. These measurements are accurate enough to extract the nuclear quadrupole moment of the $^{43}$Ca provided that the corresponding high-precision calculated electric field gradients of these states are accessible.

Previously, several values of the nuclear quadrupole moment of $^{43}$Ca have been reported. However, significant discrepancies are found among these reported values~\cite{Grundevik1979PRL, Arnold1981HI, Aydin1982ZPA,Olsson1982ZPA,Salomonson1984ZPA,Sundholm1993JCP,Yu2004PRA,Benhelm2007PRA,Sahoo2009PRA}, which span from \(-0.0408(8)\)~b to \(-0.065(20)\)~b. Grundevik~\textit{et al.} employed the atomic-beam magnetic-resonance method to precisely measure the HFS of the \(4s4p~^3\!P_{2}\) state, and determined the nuclear quadrupole moment $Q=-0.065(20)$~b~\cite{Grundevik1979PRL}.  Arnold~\textit{et al.} precisely measured the HFS of the $4s4p$ $~^3\!P_1$ state by laser and radio-frequency spectroscopy~\cite{Arnold1981HI}. Subsequently, Olsson and Salomonson reanalyzed these two measurements by taking into account the second-order correction arising from the off-diagonal hyperfine interaction among the $4s4p$$~^3\!P_J$ fine-structure levels~\cite{Olsson1982ZPA}. This reanalysis updated HFS constants for the \(4s4p~^3\!P_{1}\) and \(4s4p~^3\!P_{2}\) states and yielded a more accurate nuclear quadrupole moment of $^{43}$Ca, $Q=-0.049(5)$~b.
 Aydin \textit{et al.} applied the atomic-beam magnetic-resonance method to precisely measure the HFS of the \(4s3d~^1\!D_2\) state, obtaining a nuclear quadrupole moment of the $^{43}$Ca, $Q=-0.062(12)$~b~\cite{Aydin1982ZPA}. Salomonson used the many-body perturbation theory to evaluate HFS parameters of the \(4s4p~^3\!P_{1}\), \(4s4p~^3\!P_{2}\) and \(4s3d~^1\!D_2\) states~\cite{Salomonson1984ZPA}. By integrating the three measurements~\cite{Grundevik1979PRL, Arnold1981HI,Aydin1982ZPA}, Salomonson recommended the nuclear quadrupole moment of $^{43}$Ca as $Q=-0.049(5)$~b where the uncertainty is attributed to the theoretical scenario. A decade later, Sundholm and Olsen performed a finite element multiconfiguration Hartree-Fock calculation of electric field gradients of the \(4s3d~^1\!D_2\) state, and determined $Q$($^{43}$Ca) to be $-0.0408(8)$~b~\cite{Sundholm1993JCP}. This value is the currently adopted value~\cite{Pekka2018MP}. However, it is approximately 20\% smaller than $-0.049(5)$~b.  Benhelm ~\textit{et al.} employed the laser spectroscopy method to accurately determine the HFS constants of the $3d_{5/2}$ state in $^{43}$Ca$^{+}$~\cite{Benhelm2007PRA}. Subsequently, Sahoo adopted the relativistic coupled-cluster method to calculate the hyperfine interaction parameters of the $3d_{5/2}$ state, and
determined the nuclear quadrupole moment of $^{43}$Ca with an accuracy of 1\%~\cite{Sahoo2009PRA}. The obtained result, $Q=-0.0444(6)$~b, is approximately 8\% larger than the currently adopted value $Q=-0.0408(8)$~b~\cite{Sundholm1993JCP,Pekka2018MP}.  There are such significant differences among these reported nuclear
electric quadrupole moments of $^{43}$Ca, so it is worthwhile and essential to reinvestigate this issue.

Considering that the measurement precision of the above states is sufficiently high, performing high-precision calculations of the electric field gradients is the decisive factor in accurately obtaining the nuclear quadrupole moment of $^{43}$Ca. Accurate calculation of the electric field gradient needs to take into account both the relativistic effect and electron correlation effect. The relativistic effect can be included by solving the Dirac-Fock (DF) equation. Therefore, the electron correlation effect is the decisive factor in achieving precise values of the electric field gradient. For neutral Ca atom, the electron correlations includes core-core, core-valence, and valence-valence correlations. The three kinds of correlations are important for accurately calculating hyperfine interaction parameters. To obtain accurate and reliable electric field gradients \(q\) for the \(4s4p~^3\!P_{1}\), \(4s4p~^3\!P_{2}\) and \(4s3d~^1\!D_2\) states in $^{43}$Ca atom, we developed a comprehensive code for accurately calculating the atomic structure properties of divalent atomic systems. This is a code based on a hybrid method that combines the configuration interaction method and the coupled-cluster method. This hybrid method can comprehensively consider the core-core, core-valence, and valence-valence correlation effects simultaneously. To comprehensively evaluate the accuracy of this hybrid method, we also calculated the energies and magnetic-dipole HFS constants and compared them with available theoretical and experimental results.

This paper is organized as follows. The theoretical formulations of coupled-cluster method and configuration interaction method as well as hyperfine interaction are given in section~\ref{theory}.
Numerical results and discussions are presented in section~\ref{results}, together with
comparisons with available experimental and theoretical data.
Finally, a summary is given in section~\ref{conclusions}. Atomic units are used
throughout unless otherwise stated.

\section{Method}\label{theory}
In a many-electron atomic system, electrons are typically categorized into core electrons and valence electrons. Consequently, electron-electron correlations encompass core-core, core-valence, and valence-valence correlations. In the present work, we used the relativistic configuration interaction plus coupled-cluster method (RCICC), in which a so called
correlation potential is uesd. This correlation potential is built through a coupled-cluster (CC) calculation to depict the core-core and core-valence correlations. Meanwhile, the valence-valence correlation is accounted for through a configuration interaction (CI) calculation. First, we do a Dirac-Fock (DF) calculation on the closed-shell part to generate single-particle orbitals. These single-particle orbitals are then utilized to build the model space for CC and CI calculations. Subsequently, a CC calculation is carried out to construct the one-body and two-body correlation potentials. After that, the wave functions and energies of the system are obtained through a CI calculation with the potentials accounting for the core-core and core-valence correlations. Finally, the obtained wave functions and energies are used to evaluate different atomic properties.

The relativistic configuration interaction plus linear version of coupled-cluster theory (called RCI+all-order method) was first developed by Safronova \textit{et al.}~\cite{Safronova2009PRA}. And later, a similar method
was independently developed by Dzuba~\cite{Dzuba2014PRA}. These two methods hold the same general ideas. The method adopted in the present work is conceptually similar with the above two method except a few differences. Firstly, when constructing the correlation potential, we include not only the linear part but also the nonlinear part of the single and double excitation of cluster operators. According to the previous calculations of the properties of monovalent atomic systems~\cite{Tang2017PRA,Tang2018PRA,Tang2019JPB,Li2021pra,li2021ab,Li2021JPB}, the nonlinear terms are crucial for the energy and hyperfine interaction properties. Secondly, when calculating the transition matrix elements, we consider the random phase approximation (RPA), core Brueckner, structural radiation, and normalization corrections to all order. We also take account of the two-particle (TP) interaction to second order. In addition, we independently developed the corresponding code for accurately calculating the atomic structure properties of divalent atomic systems.

\subsection{Coupled-cluster calculation}
The exact wave function$|\Psi\rangle$ of a system can be generated when a normally-ordered wave operator $\Omega$ acts on the reference state, namely
\begin{equation}\label{1}
|\Psi\rangle=\Omega|\Phi\rangle.
\end{equation}
In the present work, the reference state $|\Phi\rangle$ is defined as the zero-order DF wave function. Within the coupled-cluster theory framework~\cite{Bartlett2007RMP}, the wave operator is expressed as the exponential of the cluster operator $S$
\begin{equation}\label{2}
\Omega = e^{S}.
\end{equation}
The cluster operator $S$ is defined in relation to a closed-shell reference determinant. Based on the number of valence holes ($m$) and the number of valence particles ($n$) to be excited relative to the reference determinant~\cite{Ilyabaev1992JCP,Ilyabaev1992CPL,Eliav1998JCP}, the cluster operator $S$ can be partitioned as follows:
\begin{align}\label{3}
S=\sum_{m\geq0}\sum_{n\geq0}\left(\sum_{\ell>m + n}S^{(m,n)}_{\ell}\right),
\end{align}
where $\ell$ denotes the number of excited electrons.

The coupled equations for the cluster operators are derived from the generalized Bloch equation by taking into account only the connected terms~\cite{Lindgren}:
\begin{align}\label{4}
{Q}[S^{(m,n)},H_{0}]{P}={Q}\left\{(V\Omega)-\chi{W}\right\}_{\rm conn}{P},
\end{align}
\begin{equation}
W = {P}V\Omega{P},
\end{equation}
where $H_{0}$ and $V$ are the zero-order DF Hamiltonian and the residual interaction, respectively. $\chi=\Omega - 1$. $W$ is the folded operator accountable for the correlation energy of the valence state, and ${P}$ and ${Q}$ are the common projection operators which act on the model space and its orthogonal complement, respectively.
In practice, the equations for the sector $S^{(0,0)}$ are first solved iteratively until convergence is achieved. Subsequently, the sector $S^{(0,1)}$ or $S^{(1,0)}$ is solved using the known $S^{(0,0)}$, and the process continues in this way. In the present work, we adopt $(m,n)=(0,0)$, $(0,1)$, and $(0,2)$, and $\ell$ is truncated at $2$, which corresponds to single and double excitations. This is the standard coupled-cluster single-double excitation (CCSD) calculation process. Previously, we have independently developed a CCSD code based on the B-splines basis set and Gauss basis set, and applied it to calculate the energies, transition matrix elements, polarizabilities, and HFS constants of monovalent atomic systems~\cite{Tang2017PRA,Tang2018PRA,Tang2019JPB,Li2021pra,li2021ab,Li2021JPB}.

 In the present work, CCSD calculations are utilized to construct the correlation potentials that characterize core-core and core-valence correlations. As a result, it is necessary to modify the coupled equations for the cluster operators.  We adopted the same scheme as Safronova and Dzuba~\cite{Safronova2009PRA,Dzuba2014PRA}, modifying the energy factor on the left-hand side of Eq.(\ref{4}) and eliminating the terms on the right-hand side of Eq.(\ref{4}) that are repeatedly accounted for in the subsequent CI calculation. Specifically:
(1) the coupled equation for $S^{(0,0)}$ remains the same as that in the standard CCSD calculation;
(2) for other cluster operators, the factor on the left-hand side of Eq.(\ref{4}) is changed from
\begin{align}\label{5}
    \begin{cases}
        (\varepsilon_{v}-\varepsilon_{r})S^{(0,1)}_{1}(rv)\\
        (\varepsilon_{v}+\varepsilon_{a}-\varepsilon_{r}-\varepsilon_{s})S^{(0,1)}_{2}(rs,va)\\
        (\varepsilon_{v}+\varepsilon_{w}-\varepsilon_{r}-\varepsilon_{s})S^{(0,2)}_{2}(rs,vw)
    \end{cases}
\end{align}
to
\begin{align}\label{6}
    \begin{cases}
        (\tilde{\varepsilon}_{v}-\varepsilon_{r})S^{(0,1)}_{1}(rv)\\
        (\tilde{\varepsilon}_{v}+\varepsilon_{a}-\varepsilon_{r}-\varepsilon_{s})S^{(0,1)}_{2}(rs,va)\\
        (\tilde{\varepsilon}_{v}+\tilde{\varepsilon}_{w}-\varepsilon_{r}-\varepsilon_{s})S^{(0,2)}_{2}(rs,vw)
    \end{cases},
\end{align}
where $a$ represents a core orbital, $r$ and $s$ are designated as virtual orbitals, and $v$ and $w$ denote valence orbitals. $\varepsilon$ is the single-particle energy, which is set as the DF energy. $\tilde{\varepsilon}$ is a pre-set energy parameter, typically selected to be the DF energy of the lowest valence state of a given symmetry. For all valence orbitals belonging to a given symmetry, this energy parameter remains the same. For example, for the Ca atom, \(\tilde{\varepsilon}(s_{1/2})=\varepsilon_{4s_{1/2}}\), \(\tilde{\varepsilon}(p_{1/2})=\varepsilon_{4p_{1/2}}\), \(\tilde{\varepsilon}(p_{3/2})=\varepsilon_{4p_{3/2}}\), and so on. Meanwhile, \(S^{(0,1)}_{1}\), \(S^{(0,2)}_{2}\), and the folded operator $W$ on the right-hand side of Eq.(\ref{4}) are removed.
\begin{widetext}
\subsection{Configuration Interaction Calculation}
In a divalent atomic system, the equation for the effective interaction can be formulated as
\begin{equation}\label{7}
    \left(\sum_{i = 1}^{2}H_{1}(r_{i})+V_{2}(r_{12})\right)\vert\Psi(\pi JM)\rangle = E\vert\Psi(\pi JM)\rangle,
\end{equation}
where \(H_{1}\) and \(V_{2}\) denote the one-body and two-body interaction Hamiltonians, respectively. The one-body Hamiltonian is given by
\begin{equation}\label{8}
    H_{1}=H_{\rm DF}+\Sigma_{1},
\end{equation}
where \(H_{\rm DF}\) is the DF Hamiltonian, and \(\Sigma_{1}\) corresponds to the one-body correlation potential.
The two-body interaction Hamiltonian is expressed as
\begin{equation}\label{9}
    V_{2}=\frac{1}{r_{12}}+\Sigma_{2},
\end{equation}
where the first term represents the electron-electron Coulomb interaction, while the second term is the two-body correlation potential.
The wave function \(\vert\Psi(\pi JM)\rangle\) of the system is described as a linear combination of configuration wave functions sharing the same parity \(\pi\), angular momentum \(J\), and magnetic quantum number \(M\), and
\begin{equation}\label{10}
    \vert\Psi(\pi JM)\rangle=\sum_{v\leq w}C_{vw}\vert\Phi_{vw}(\pi JM)\rangle,
\end{equation}
where \(C_{vw}\) are the expansion coefficients. The configuration wave function is constructed from single-particle orbitals:
\begin{equation}\label{11}
    \vert\Phi_{vw}(\pi JM)\rangle=\eta_{vw}\sum_{m_{v},m_{w}}\langle j_{v}m_{v},j_{w}m_{w}\vert JM\rangle a_{v}^{\dagger}a_{w}^{\dagger}\vert0\rangle.
\end{equation}
The symmetry factor \(\eta_{vw}\) is defined as:
\begin{equation}\label{12}
    \eta_{vw}=\begin{cases}
        \frac{\sqrt{2}}{2},& v = w\\
        1,& v\neq w
    \end{cases}.
\end{equation}
The configuration wave function $\vert\Phi_{vw}\rangle$ is an eigenstate of \(H_{\rm DF}\), with energy \(\varepsilon_{v}+\varepsilon_{w}\). By substituting Eq.(\ref{10}) into Eq.~(\ref{7}) and applying the variational principle, a general eigenvalue equation can be derived
\begin{equation}\label{13}
    \sum_{x < y}[(H_{1})_{vw,xy}+(V_{2})_{vw,xy}]C_{xy}=EC_{vw}.
\end{equation}

The matrix elements of the one-body interaction Hamiltonian are
\begin{align}\label{14}
    (H_{1})_{vw,xy}=(\varepsilon_{v}+\varepsilon_{w})\delta_{vx}\delta_{wy}+\eta_{vw}\eta_{xy}
    \times\bigg((\Sigma_{1})_{vx}\delta_{wy}+(\Sigma_{1})_{wy}\delta_{vx}
    +(-1)^{J}\left((\Sigma_{1})_{vy}\delta_{wx}+(\Sigma_{1})_{wx}\delta_{vy}\right)\bigg).
\end{align}

The matrix elements of the two-body interaction Hamiltonian are
\begin{small}
\begin{align}\label{15}
    (V_{2})_{vw,xy}=\eta_{vw}\eta_{xy}
    \left\{C_{1}\begin{Bmatrix}j_{v}&j_{w}&J\\j_{y}&j_{x}&L\end{Bmatrix}\bigg(X_{L}(vw,xy)+(\Sigma_{2})_{L}(vw,xy)\bigg)
    +C_{2}\begin{Bmatrix}j_{v}&j_{w}&J\\j_{x}&j_{y}&L\end{Bmatrix}\bigg(X_{L}(vw,yx)+(\Sigma_{2})_{L}(vw,yx)\bigg)
    \right\},
\end{align}
\end{small}
with
\begin{small}
\begin{align}\label{16}
\begin{cases}
C_{1}=(-1)^{(J + L+j_w+j_x)}\\
C_{2}=(-1)^{(L+j_w+j_x)}\\
    X_{L}(vw,xy)=(-1)^{L}\langle\kappa_v\parallel C^{L}\parallel\kappa_x\rangle\langle\kappa_w\parallel C^{L}\parallel\kappa_y\rangle R_{L}(vw,xy)
\end{cases}.
\end{align}
\end{small}
In the above expressions, $R_{L}(vw,xy)$ and $\langle\kappa_v\parallel C^{L}\parallel\kappa_x\rangle$ represent the two-electron integral and the angular reduced matrix element , respectively. They are defined as:
\begin{align}\label{17}
    R_{L}(vw,xy)&=\int_{0}^{\infty}\bigg(f_{v}(r_{1})f_{x}(r_{1})+g_{v}(r_{1})g_{x}(r_{1})\bigg)dr_{1}
    \int_{0}^{\infty}\frac{r_{<}^{L}}{r_{>}^{L + 1}}\bigg(f_{w}(r_{2})f_{y}(r_{2})+g_{w}(r_{2})g_{y}(r_{2})\bigg)dr_{2},
\end{align}
and
\begin{align}\label{18}
    \langle\kappa_v\parallel C^{L}\parallel\kappa_x\rangle&=(-1)^{j_v+\frac{1}{2}}\sqrt{(2j_v + 1)(2j_x + 1)}
    \times
    \left(\begin{array}{ccc}
        j_v&j_x&L\\
        -\frac{1}{2}&\frac{1}{2}&0\\
    \end{array}\right)\Pi(\ell_v,L,\ell_x),
\end{align}
where \(\Pi(\ell_v,L,\ell_x)=1\) when \(\ell_v+L+\ell_x\) is even; otherwise, \(\Pi(\ell_v,L,\ell_x)=0\). The relativistic angular-momentum quantum number \(\kappa=\ell(\ell + 1)-j(j + 1)-\frac{1}{4}\). $f$ and $g$ are the large and small radial components of the Dirac wave function, respectively.

The matrix elements of the one-body and two-body correlation potentials, \((\Sigma_{1})_{xv}\) and \((\Sigma_{2})_{L}(xy,vw)\), are obtained through the coupled-cluster calculation:
\begin{equation}\label{19}
    \begin{cases}
        (\Sigma_{1})_{xv}=(\tilde{\varepsilon}_{v}-\varepsilon_{x})S_{1}^{(0,1)}(xv)\\
        (\Sigma_{2})_{L}(xy,vw)=(\tilde{\varepsilon}_{v}+\tilde{\varepsilon}_{w}-\varepsilon_{x}-\varepsilon_{y})(S_{2}^{(0,2)})_{L}(xy,vw)	
    \end{cases}.
\end{equation}
When constructing the one-body and two-body correlation potentials, we only considered the contributions of single and double excited states. To compensate for the higher-order correlation effects that were not taken into account, we introduced a rescaling parameter $\rho_{\kappa}$ and substitute $\rho_{\kappa}\Sigma_{1}$ for the one-body correlation potential $\Sigma_{1}$. By adjusting the value of the rescaling parameter, the calculated energy can be made closer to the experimental energy. This scheme has already been used in our previous RCI+MBPT calculations~\cite{Zhang2023PRA,Zhang2024PRA}. Actually these rescaling parameters are close to $1$ since the energies calculated by RCI+CCSD method show relatively very small difference from experimental values.

\subsection{Reduced Transition Matrix Element calculation}

After obtaining wave functions of the system, the reduced transition matrix
element of operator $O$ with order $k$ from state $|\Psi({\pi}JM)\rangle$ to $|\Psi({\pi'}J'M')\rangle$  can be evaluated using the following expressions~\cite{Johnson}:
\begin{small}
 \begin{align}\label{one}
\langle\Psi({\pi}J)\|O^{(k)}\|\Psi({\pi'}J')\rangle=&(-1)^{k}\sqrt{(2J+1)(2J'+1)}
\sum_{v<w,x<y}\eta_{vw}\eta_{xy}C_{vw}C_{xy}
\left\{(-1)^{j_{y}+j_{v}+J'}\begin{Bmatrix}J&J'&k\\j_{x}&j_{v}&j_{y}\end{Bmatrix}o^{(k)}_{vx}\delta_{wy}\right.\notag\\
&+\left.(-1)^{j_{y}+j_{v}}\begin{Bmatrix}J&J'&k\\j_{y}&j_{v}&j_{x}\end{Bmatrix}o^{(k)}_{vy}\delta_{wx}
+(-1)^{J+J'+1}\begin{Bmatrix}J&J'&k\\j_{x}&j_{w}&j_{y}\end{Bmatrix}o^{(k)}_{wx}\delta_{vy}+(-1)^{j_{x}+j_{y}+J}\begin{Bmatrix}J&J'&k\\j_{y}&j_{w}&j_{x}\end{Bmatrix}o^{(k)}_{wy}\delta_{vx}\right\},
\end{align}
\end{small}
where $o^{(k)}_{vx}=\langle{\Psi_{v}}\|o^{(k)}\|{\Psi_{x}}\rangle$ is the single-electron reduced matrix element.  In the standard CC calculation, the reduced matrix element of monovalent atomic system is calculated using the following formula:
\begin{small}
\begin{align}\label{QQ}
o^{(k)}_{vx}=\frac{\langle{\Psi_{v}}\|o^{(k)}\|\Psi_{x}\rangle}{\sqrt{\langle{\Psi_{v}}|\Psi_{v}\rangle}\sqrt{\langle{\Psi_{x}}|\Psi_{x}\rangle}}
=\frac{\langle{\Phi_{v}}\|e^{S\dag}o^{(k)}{e^{S}}\|\Phi_{x}\rangle}{\sqrt{\langle{\Phi_{v}}|e^{S\dag}e^{S}|\Phi_{v}\rangle}\sqrt{\langle{\Phi_{x}}|e^{S\dag}e^{S}|\Phi_{x}\rangle}}.
\end{align}
\end{small}
At the LCCSD approximation,
\begin{small}
\begin{align}\label{es}
e^{S\dag}O{e^{S}}\approx&O+\{{O}{S^{(0,0)}_{1}}+{\rm c.c.}\}+\{{O}{S^{(0,1)}_1}+{\rm c.c.}\}
                 +\{{O}{S^{(0,1)}_{2}}+{\rm c.c.}\}+\{S^{(0,0)\dag}_1O{S^{(0,1)}_1}+{\rm c.c.}\}\notag\\
                 &+S^{(0,0)\dag}_1O{S^{(0,0)}_1}+\{S^{(0,0)\dag}_1O{S^{(0,0)}_2}+{\rm c.c.}\}
                 +\{S^{(0,0)\dag}_1O{S^{(0,1)}_2}+{\rm c.c.}\}+S^{(0,0)\dag}_2O{S^{(0,0)}_2}\notag\\
                 &+\{S^{(0,0)\dag}_2O{S^{(0,1)}_2}+{\rm c.c.}\}+S^{(0,1)\dag}_1O{S^{(0,1)}_1}
                 +\{S^{(0,1)\dag}_1O{S^{(0,1)}_2}+{\rm c.c.}\}+S^{(0,1)\dag}_2O{S^{(0,1)}_2},
\end{align}
\end{small}
and
\begin{small}
\begin{align}\label{qs}
e^{S\dag}{e^{S}}\approx 1+{S^{(0,0)\dag}_1}{S^{(0,0)}_1}+{S^{(0,1)\dag}_1}{S^{(0,1)}_1}
                        +{S^{(0,0)\dag}_2}{S^{(0,0)}_2}+{S^{(0,1)\dag}_2}{S^{(0,1)}_2},
\end{align}
\end{small}
where ${\rm c.c.}$ stands for the complex conjugate part. However, in the RCI+CCSD calculations, terms involving $S^{(0,1)}_{1}$ in Eq.(\ref{es}) and Eq.(\ref{qs}) need to be removed, because these terms have been included in the CI calculation. In our previous RCI+MBPT calculations, the single-electron transition matrix elements typically only included the RPA correction terms~\cite{Zhang2023PRA,Zhang2024PRA}. In fact, it has been observed that core Brueckner, structural radiation, and normalization corrections are also important in the calculations of the hyperfine interaction properties of monovalent atomic systems~\cite{Blundell1989PRA,Tang2019JPB,Li2021pra,li2021ab,Li2021JPB}. It should be noted that Eq.(\ref{QQ}) includes the RPA, core Brueckner, structural radiation, and normalization corrections to all-order~\cite{Blundell1989PRA}.

In addition, the contribution of the two-particle interaction to the transition matrix element also needs to be considered. The TP correction is also significant for some atomic states~\cite{Porsev2022PRA}. In the present work, we consider the TP correction by the second-order many-body perturbation calculation~\cite{Safronova1999JPB,Savukov2004PRA}. The expression of the TP correction is as follows:
\begin{small}
\begin{align}\label{two}
\langle\Psi({\pi}J)\|O^{(k)}_{\rm TP}\|\Psi({\pi'}J')\rangle=&\sqrt{(2J+1)(2J'+1)}
\sum_{v<w,x<y}\eta_{vw}\eta_{xy}C_{vw}C_{xy}\times\left\{\tilde{O}^{(k)}_{vw,xy}
+(-1)^{(j_x+j_y+J'+1)}\tilde{O}^{(k)}_{vw,yx}\right.\notag\\
&+\left.(-1)^{(j_v+j_w+J+1)}\tilde{O}^{(k)}_{wv,xy}
+(-1)^{(j_x+j_y+J'+j_v+j_w+J)}\tilde{O}^{(k)}_{wv,yx}\right\},
\end{align}
\end{small}
with
\begin{small}
\begin{align}
\tilde{O}^{(k)}_{vw,xy}=&\sum_{L,a}(-1)^{k+L+j_w+j_y+J'}\begin{Bmatrix}J'&J&k\\j_{a}&j_{x}&j_{y}\end{Bmatrix}\begin{Bmatrix}J&j_a&j_y\\L&j_{v}&j_{w}\end{Bmatrix}
\times\frac{o^{(k)}_{ax}X_{L}(vw,xy)}{\varepsilon_{a}+\varepsilon_{y}-\varepsilon_{v}-\varepsilon_{w}}\notag\\
&+\sum_{L,a}(-1)^{k+L+j_{v}+j_{x}}\begin{Bmatrix}J&J'&k\\j_{a}&j_{v}&j_{w}\end{Bmatrix}\begin{Bmatrix}J'&j_a&j_w\\L&j_{y}&j_{x}\end{Bmatrix}
\times\frac{o^{(k)}_{va}X_{L}(aw,xy)}{\varepsilon_{a}+\varepsilon_{w}-\varepsilon_{x}-\varepsilon_{y}}.
\end{align}
\end{small}
Therefore, the reduced transition matrix element is the sum of Eq.(\ref{one}) and Eq.(\ref{two}). In the present work, the operator $o$ is hyperfine interaction operator.

\subsection{Hyperfine-structure Constant}
The hyperfine-structure of the atomic energy level results from the interaction between electrons and the electromagnetic multipole moments of the nucleus. In comparison to fine-structure splitting, hyperfine splitting is smaller. Consequently, the hyperfine interaction can be regarded as a perturbation. When only considering the first-order corrections, the hyperfine energy can be parameterized as follows:
\begin{small}
\begin{align}
\Delta{E_F^{(1)}}=&\frac{A}{2}K+\frac{B}{2}\frac{3K(K+1)-4I(I+1)J(J+1)}{2I(2I-1)2J(2J-1)},
\end{align}
\end{small}
where $K=F(F+1)-I(I+1)-J(J+1)$,  $A$ and $B$ are the magnetic dipole and the electric quadrupole HFS constant, which are defined as\cite{li2021ab}:
\begin{small}
\begin{align}
A=\frac{\mu}{I}\frac{\langle\gamma J\|T^{(1)}\|\gamma J\rangle}{\sqrt{J(J+1)(2J+1)}},
\end{align}
\end{small}
and
\begin{small}
\begin{align}
B=2Q\bigg[\frac{2J(2J-1)}{(2J+1)(2J+2)(2J+3)}\bigg]^{1/2}\langle\gamma J\|T^{(2)}\|\gamma J\rangle,
\end{align}
\end{small}
respectively, where $\gamma$ represents the quantum numbers besides $J$, and $T^{(k)}=\sum_{i}{t^{(k)}(\textbf{r}_{i})}$. The single-particle reduced matrix elements of the operators $t^{(1)}$ and $t^{(2)}$ are given by:
\begin{small}
\begin{align}
\langle{\kappa_{a}}\|t^{(1)}\|\kappa_{b}\rangle=-(\kappa_{a}+\kappa_{b})\langle-\kappa_{a}\|C^{(1)}\|\kappa_{b}\rangle
\int_{0}^{\infty}{\frac{f_{a}(r)g_{b}(r)+f_{b}(r)g_{a}(r)}{r^2}\times{F^{(1)}(r)}dr},
\end{align}
\end{small}
and
\begin{small}
\begin{align}
\langle{\kappa_{a}}\|t^{(2)}\|\kappa_{b}\rangle=-\langle\kappa_{a}\|C^{(2)}\|\kappa_{b}\rangle\int_{0}^{\infty}{\frac{f_{a}(r)f_{b}(r)+g_{a}(r)g_{b}(r)}{r^3}\times{F^{(2)}(r)}dr}.
\end{align}
\end{small}
\end{widetext}
Here, the nuclear distribution function $F^{(k)}(r)$ is defined as:
\begin{align}
	F^{(k)}(r) =\begin{cases}
		(\frac{r}{R_{N}})^{2k+1},&r\leq{R_{N}}\\
		1,&r>R_{N}
	\end{cases},
\end{align}
where $R_{N}=\sqrt{5/3}\langle{r^{2}}\rangle^{1/2}$ is the radius of the sphere, and $\langle{r^{2}}\rangle^{1/2}$ is the charge root-mean-square radius of the nucleus.

Based on above definitions of the HFS constants, the nuclear electric quadrupole moment $Q$ (in b) can be extracted from the experimental values of the HFS constant $B$
by
\begin{align}
Q=\frac{B}{234.9648867q},
\end{align}
where the HFS constant $B$ is in the unit of MHz, and the electric field gradient $q$ is defined as
\begin{small}
\begin{align}\label{eq9}
{q}=2\bigg[\frac{J(2J - 1)}{(2J + 1)(J + 1)(2J + 3)}\bigg]^{1/2}\langle\gamma J\|T^{(2)}\|\gamma J\rangle.
\end{align}
\end{small}
that is expressed in atomic units.
\subsection{Computation details}

Similar to Refs.~\cite{Li2021pra,Li2021JPB}, the large and small components of the Dirac wave functions are expanded using a finite basis set composed of even-tempered Gaussian-type functions~\cite{Chaudhuri1999pra}. The Gaussian-type function has the form:
 \begin{align}
G_{i,\kappa}={\aleph}_{i}r^{n_{\kappa}}e^{-\alpha_{i}{r^2}},
\end{align}
where $\aleph_{i}$ is the normalization factor, $n_{\kappa}=\ell+1$, and $\alpha_{i}=\alpha\beta^{i-1}$. To avoid the spurious state and variational collapse problem, the large and small components need to satisfy "kinetically-balanced" condition, i.e.,
\begin{align}
\left\{\begin{array}{l}
f_{\kappa}(r)=\sum_{i=1}^{N}{C^{f_\kappa}_{i}G_{i,\kappa}(r)}\\
g_{\kappa}(r)=\sum_{i=1}^{N}{C^{g_\kappa}_{i}(\frac{d}{dr}+\frac{\kappa}{r})G_{i,\kappa}(r)}
\end{array}\right ..
\end{align}
\begin{table}
\caption{The parameters of the Gauss basis set. N is the number of basis set for each symmetry.}
\label{tab1}
\begin{ruledtabular}
\begin{tabular}{ccccccccccc}
 &s&p&d&f&g&h&i\\
\hline
$\alpha$&0.00085&0.00085&0.00085&0.0026&0.086&0.086&0.086\\
$\beta$&1.88&1.87&1.89&1.91&2.0&2.0&2.0\\
N    &35&30&30&25&15&15&15\\
\end{tabular}
\end{ruledtabular}
\end{table}

Table~\ref{tab1} lists the parameters of Gauss basis set used in the present work. In DF calculation, the Fermi nuclear distribution is used to describe the Coulomb potential between electrons and the nucleus.  In CC calculation, the $n(4-7)s_{1/2}$, $n(4-7)p_{1/2,3/2}$, and $n(3-6)d_{3/2,5/2}$ are set as the valence orbitals, the single-particle orbitals with energy smaller than 20000 $a.u.$ are set as the virtual orbitals, and the partial wave $\ell_{max}$ is limited to $6$.
In CI calculation, the single-particle orbitals with energy smaller than 500 $a.u.$ are used to construct configuration, and the partial wave $\ell_{max}$ is limited to 4.
In the second-order many-body perturbation calculations, the summation is carried out over the entire basis set.

To assess the influence of electron correlation effects in the computation of the energy and hyperfine interaction properties of neutral calcium (Ca) atom, we adopted four distinct approaches when formulating the one-body and two-body correlation potentials.

Method 1: The one-body and two-body correlation potentials are derived using the second-order many-body perturbation theory. The detailed expressions of correlation potentials were given in Ref.~\cite{Safronova2009PRA}. This method is denoted as RCI+ MBPT(2).

Method 2: The one-body and two-body correlation potentials are constructed via linear coupled-cluster singles and doubles (LCCSD) calculations. This method is labeled as RCI+LCCSD. In Ref.~\cite{Safronova2009PRA}, it was referred to as RCI+all-order.

Method 3: The one-body and two-body correlation potentials are established through full coupled-cluster singles and doubles (CCSD) calculations. This method is designated as RCI+CCSD. In contrast to the RCI+LCCSD approach, this method accounts for the non-linear terms associated with single and double excitations of the cluster operator.

Method 4: Starting from the one-body and two-body correlation potentials obtained from CCSD calculations, the two-body correlation potentials are held constant. Meanwhile, a rescaling parameter is applied to the one-body potential. This approach is marked as RCI+CCSDs. The specific rescaling parameters are $\rho_{-1}=0.981$, $\rho_{1}=1.015$, $\rho_{-2}=1.015$, $\rho_{2}=1.035$, $\rho_{-3}=1.035$, and $\rho_{\rm others}=1.0$.

\section{Results and Discussion}\label{results}

\subsection{The energy of the low-lying states in Neutral Ca atom}

Table~\ref{Tab2} presents the energies of the atomic states in the $4s^2$, $4s4p$, $4s3d$, $4s5s$, $4s5p$, and $4s4d$ configurations of the neutral Ca atom. These energies are obtained by using RCI+MBPT(2), RCI+LCCSD, RCI+CCSD, and RCI+CCSDs methods. To present the data in a more organized and accessible table format, RCI+MBPT(2), RCI+LCCSD, RCI+CCSD, and RCI+CCSDs are abbreviated as $\rm M_{1}$, $\rm M_{2}$, $\rm M_{3}$, and $\rm M_{4}$ respectively. We also compare our calculated results with the results obtained by the RCI+all-order method~\cite{Safronova2009PRA}, as well as the experimental values from National
Institute of Standards and Technology (NIST)~\cite{NIST_ASD}. The RCI+all-order method is denoted as $\rm M_{5}$. The symbol $\delta_{n}$ represents the relative difference between the theoretical results obtained by the $\rm M_{n}$ method and the experimental values, and is given in percentages.

Table~\ref{Tab2} reveals that the RCI+MBPT(2) method yields the most significant disparity when compared with the experimental values, with a relative difference spanning from 0.5\% to 1.5\%. The RCI+LCCSD method shows a difference ranging from 0.2\% to 0.6\%, while the RCI+CCSD method exhibits a difference of less than 0.3\%. Evidently, the RCI+CCSD method surpasses the RCI+LCCSD method in terms of accuracy, achieving at least a two-fold improvement. This finding indicates the crucial role of nonlinear terms in determining energy properties, aligning well with the observations in monovalent atomic systems~\cite{Tang2017PRA,Tang2019JPB}. The RCI+CCSDs showcases a remarkable enhancement in accuracy, with a difference of less than 0.2\% from the experimental results. Notably, our RCI+LCCSD results are very close to the RCI+all-order results reported by Safronova et al~\cite{Safronova2009PRA}. For the $4s^{2}$ and $4s4p$ configurations, the difference between the two sets of results is less than 50 cm$^{-1}$. However, a divergence of 220 cm$^{-1}$ is observed for the $4s3d$ configuration. This divergence can potentially be attributed to the subtleties in the construction of one-body and two-body correlation potentials. The $3d$ orbital of Ca$^{+}$, which is more sensitive to electron correlation effects compared to the $4s$ and $4p$ orbitals of Ca$^{+}$, likely accounts for this difference.

\begin{table*}
\caption[]{The energies (in cm$^{-1}$) of the atomic states in the $4s^2$, $4s4p$, $4s3d$, $4s5s$, $4s5p$, and $4s4d$ configurations of the neutral Ca atom, calculated in RCI+MBPT(2), RCI+LCCSD, RCI+CCSD, and RCI+CCSDs approximation, are presented. The RCI RCI+MBPT(2), RCI+LCCSD, RCI+CCSD, and RCI+CCSDs methods are denoted as $\rm M_{1}$, $\rm M_{2}$, $\rm M_{3}$, and $\rm M_{4}$, respectively. $\rm M_{5}$ refers to RCI+all-order method presented in Ref.~\cite{Safronova2009PRA}. The experimental values are taken from  NIST~\cite{NIST_ASD}.
$\delta_{n}$ represents the relative difference between the theoretical results obtained by the $\rm M_{n}$ method and the experimental values, and this relative difference is given in percentages.}\label{Tab2}
	\begin{ruledtabular}
		\begin{tabular}{cccccccccccccccc}
Conf.    &Terms& $\rm M_{1}$ & $\rm M_{2}$& $\rm M_{3}$& $\rm M_{4}$ & $\rm M_{5}$~\cite{Safronova2009PRA}&NIST~\cite{NIST_ASD} & $\delta_{1}$ & $\delta_{2}$& $\delta_{3}$& $\delta_{4}$& $\delta_{5}$ \\
\hline
$4s^2$&$^1\!S_{0}$  &$-146049$&$-145565$&$-145233$&$-145107   $&$-145517$&$-145058$&0.68&	0.35&	0.12&	0.03 &0.32\\
4s4p &$^3\!P_{0}$	&$-130547$&$-130202$&$-129945$&$-129905   $&$-130179$&$-129900$&0.50&	0.23&	0.03&	0.01 &0.21\\
4s4p &$^3\!P_{1}$	&$-130496$&$-130156$&$-129899$&$-129859   $&$-130132$&$-129848$&0.50&	0.24&	0.04&	0.01 &0.22\\
4s4p &$^3\!P_{2}$	&$-130388$&$-130044$&$-129787$&$-129747   $&$-130019$&$-129742$&0.50&	0.23&	0.03&	0.01 &0.21\\
4s4p &$^1\!P_{1}$	&$-122282$&$-121832$&$-121471$&$-121449   $&$-121788$&$-121405$&0.72&	0.35&	0.05&	0.04 &0.32\\
4s3d &$^3\!D_{1}$   &$-126343$&$-125409$&$-124463$&$-124668   $&$-125182$&$-124722$&1.30&	0.55&	0.21&	0.04 &0.37\\
4s3d &$^3\!D_{2}$	&$-126325$&$-125391$&$-124447$&$-124651   $&$-125162$&$-124709$&1.30&	0.55&	0.21&	0.05 &0.36\\
4s3d &$^1\!D_{2}$	&$-124637$&$-123770$&$-122941$&$-123134   $&$-123552$&$-123208$&1.16&	0.46&	0.22&	0.06 &0.28\\
4s5s &$^1\!S_{0}$	&$-112475$&$-112067$&$-111822$&$-111732   $&$-112051$&$-111741$&0.66&	0.29&	0.07&	0.01 &0.27\\
4s5s &$^3\!S_{1}$	&$-114257$&$-113829$&$-113590$&$-113498   $&$-113823$&$-113518$&0.65&	0.27&	0.06&	0.02 &0.28\\
4s5p &$^3\!P_{0}$	&$-109324$&$-108847$&$-108557$&$-108513   $&         &$-108510$&0.75&	0.31&	0.04&	0.01 \\
4s5p &$^3\!P_{1}$	&$-109318$&$-108841$&$-108551$&$-108506   $&         &$-108503$&0.75&	0.31&	0.04&	0.01 \\
4s5p &$^3\!P_{2}$	&$-109297$&$-108820$&$-108529$&$-108485   $&         &$-108483$&0.75&	0.31&	0.04&	0.01 \\
4s5p &$^1\!P_{1}$	&$-109171$&$-108672$&$-108353$&$-108330   $&         &$-108326$&0.78&	0.32&	0.02&	0.01 \\
4s4d &$^3\!D_{1}$	&$-108096$&$-107639$&$-107220$&$-107164   $&         &$-107310$&0.73&	0.31&	0.08&	0.14 \\
4s4d &$^3\!D_{2}$	&$-108639$&$-108115$&$-107635$&$-107635   $&         &$-107760$&0.82&	0.33&	0.12&	0.12 \\
4s4d &$^1\!D_{2}$	&$-108092$&$-107635$&$-107217$&$-107161   $&         &$-107306$&0.73&	0.31&	0.08&	0.14 \\
		\end{tabular}
	\end{ruledtabular}
\end{table*}

\subsection{Magnetic dipole hyperfine-structure constant $A$ }

Table~\ref{Tab3} lists the magnetic dipole hyperfine-structure constants of 4s3d $^1\!D_{2}$, 4s4p $^3\!P_{1}$, and 4s4p $^3\!P_{2}$ states in $^{43}$Ca atom and compares them with other theoretical and experimental results. The magnetic moment of $^{43}$Ca used here \((I=7/2,\mu=-1.317643)\) is taken from Ref.~\cite{Stone2005ADNDT}. The root-mean-square radius of the nucleus used here ($\langle{r^{2}}\rangle^{1/2}=3.4954$~fm) is from Ref.~\cite{Angeli2013ADNDT}.
Some previous experimental works have reported the HFS constants $A$ of some states~\cite{Grundevik1979PRL,Arnold1981HI,Aydin1982ZPA}. In this table, we only list the most accurate measured results. Similar to the case of energy properties, we also list the calculated values obtained by four methods: RCI+MBPT(2), RCI+LCCSD, RCI+CCSD, and RCI+CCSDs. From Table~\ref{Tab3}, it can be observed that the results of the RCI+MBPT method have the largest difference from the experimental values, being 14\%, 6\%, and 6.6\% for 4s3d $^1\!D_{2}$, 4s4p $^3\!P_{1}$, and 4s4p $^3\!P_{2}$ states. The results calculated by the other three methods are very close to each other, with the maximum difference not exceeding 2\%. The results of RCI+LCCSD, RCI+CCSD, and RCI+CCSDs methods are also very close to the experimental values, and the differences between them and the experimental values are all less than 2\%. In our previous works~\cite{Tang2019JPB,Li2021pra,li2021ab,Li2021JPB}, we found that the CCSD method may be superior to the LCCSD method for calculating the hyperfine-structure constants of monovalent atomic systems. However, we find that the result of RCI+LCCSD method is the closest to the experimental result for Ca atom. It implies that the higher-order correlation effects beyond CCSD may have the opposite sign compared to the contributions of the nonlinear terms of single and double clusters, and they will cancel each other out. Therefore, we take the value of RCI+LCCSD as the final value, and the maximum difference between this value and the results of RCI+CCSD or RCI+CCSDs is taken as the uncertainty. This way will be applied to the electric field gradients $q$ of 4s3d $^1\!D_{2}$, 4s4p $^3\!P_{1}$, and 4s4p $^3\!P_{2}$ states in $^{43}$Ca atom.
We also compare other theoretical results~\cite{Beloy2008PRA,Porsev2004PRA}. The results reported in Ref.~\cite{Beloy2008PRA} were obtained using the RCI + MBPT(2) method, which is identical to the RCI+MBPT(2) method employed in the present work. The discrepancy in the results can be attributed to the fact that our calculation of the transition matrix elements incorporates the contribution of the TP interaction correction. If only the RPA correction is included, our RCI+MBPT(2) result for 4s4p $^3\!P_{2}$ state, $-181.5$ MHz, is very close to the result, $-179.9$ MHz, in the Ref.~\cite{Beloy2008PRA}.  Our final results are in agreement with the result from RCI+MBPT(2) method by Porsev et al~\cite{Porsev2004PRA}. Their RCI+MBPT(2) method has a slight difference from ours. Their HFS constants include RPA correction and other corrections.
\begin{table}
	\caption[ ]{Hyperfine-structure constant $A$ (in MHz) of 4s3d $^1\!D_{2}$, 4s4p $^3\!P_{1}$, and 4s4p $^3\!P_{2}$ states in $^{43}$Ca \(( I=7/2,\mu=-1.317643 )\) atom. }\label{Tab3}
	\begin{ruledtabular}
		\begin{tabular}{clllllllllll}
Method     &4s3d $^1\!D_{2}$&4s4p $^3\!P_{1}$&4s4p $^3\!P_{2}$\\
\hline
RCI+MBPT(2)&$-15.21$	   &$-211.3$	&$-183.4$\\
RCI+LCCSD  &$-17.44$	   &$-198.2$	&$-171.7$\\
RCI+CCSD   &$-17.37$	   &$-196.9$	&$-170.7$\\
RCI+CCSDs  &$-17.54$	   &$-196.3$	&$-170.1$\\
Final result    &$-17.4(2)$     &$-198(2)$	&$-172(2)$\\
RCI+MBPT~\cite{Beloy2008PRA}      &     &            &$-179.9$\\
RCI+MBPT~\cite{Porsev2004PRA}     &      &$-199.2$     &$-173.1$\\
Expt.      &$-17.650(2)$	  &$-198.871(2)$	&$-171.959(2)$\\
           &~\cite{Aydin1982ZPA}&~\cite{Arnold1981HI,Olsson1982ZPA}&~\cite{Grundevik1979PRL,Olsson1982ZPA} \\
\end{tabular}
\end{ruledtabular}
\end{table}

\subsection{Nuclear electric quadrupole moment $Q$}

Table~\ref{Tab4} presents the electric field gradients $q$ (in a.u.) of 4s3d \(^1\!D_{2}\), 4s4p \(^3\!P_{1}\), and 4s4p \(^3\!P_{2}\) states in the \(^{43}\)Ca atom.
Similar to the HFS constants \(A\), we list the results from four methods, and give the final values and corresponding uncertainties. For the 4s3d \(^1\!D_{2}\) state, the results obtained by the four methods are relatively close, with a difference of less than 3\%. This indicates that for the 4s3d \(^1\!D_{2}\) state, there is a mutual cancellation among the contributions from high-order correlations beyond MBPT(2). However, for the \(4s4p~^3\!P_{1,2}\) state, there is a difference of about 7\% between the result of RCI+MBPT(2) and those of the other three methods. For this configuration, the electron correlation effects beyond MBPT(2) are very important, which is similar to the case of the HFS constant \(A\). The results calculated by RCI+LCCSD, RCI+CCSD, and RCI+CCSDs are very close to each other, with the maximum difference not exceeding 2\%. As in the case of HFS constant \(A\), the result of RCI+LCCSD is taken as the final result, and the maximum difference between RCI+LCCSD results and those of RCI+CCSD and RCI+CCSDs is taken as the uncertainty.

\begin{table}
	\caption[ ]{The electric field gradients $q$ (in a.u.) of 4s3d $^1\!D_{2}$, 4s4p $^3\!P_{1}$, and 4s4p $^3\!P_{2}$ states in $^{43}$Ca atom.}\label{Tab4}
	\begin{ruledtabular}
		\begin{tabular}{cccccccccccc}
Method      &4s3d $^1\!D_{2}$&4s4p $^3\!P_{1}$&4s4p $^3\!P_{2}$\\
\hline
RCI+MBPT(2) &	0.4203 	   &	-0.2589 	 &	0.5086 	\\
RCI+LCCSD	&	0.4113 	   &	-0.2377 	 &	0.4676 	\\
RCI+CCSD	&	0.4157     &	-0.2339 	 &	0.4602 	\\
RCI+CCSDs	&	0.4187 	   &	-0.2347 	 &	0.4617 	\\
Final result     &   0.411(8)   &    -0.238(4)    &  0.468(8) \\
\end{tabular}
\end{ruledtabular}
\end{table}

Combining our calculated electric field gradient \(q\) with the experimental values \(B(4s4p~^3\!P_{1}) = 2.672(16)\) MHz~\cite{Arnold1981HI,Olsson1982ZPA}, \(B(4s4p~ ^3\!P_{2})=-5.275(14)\) MHz~\cite{Grundevik1979PRL,Olsson1982ZPA}, and \(B(4s3d~^1\!D_{2})=-4.642(12)\) MHz~\cite{Aydin1982ZPA}, we can obtain three results of the nuclear electric quadrupole moment \(Q\) of \(^{43}\)Ca. These results are presented in Table~\ref{Tab5} and compared with other available values~\cite{Grundevik1979PRL,Aydin1982ZPA,Salomonson1984ZPA,Olsson1982ZPA,Sundholm1993JCP,Silverans1991ZPD,Sahoo2009PRA}. As can be seen from Table~\ref{Tab5}, the values of the nuclear quadrupole moment $Q$ for the three states are very close. The measured HFS constants \(B\) of these three states have similar and sufficient accuracies. Thus, the uncertainties of our determined electric quadrupole moment \(Q\) come entirely from the theoretical aspect. Based on the nuclear quadrupole moments obtained from the three states, we obtain the final result and the corresponding uncertainty, being \(Q = - 0.0479(6)\)~b. As shown in Table~\ref{Tab5}, our final result is consistent with the results in earlier references \(- 0.049(5)\)~b~\cite{Salomonson1984ZPA,Olsson1982ZPA}, but our uncertainty is smaller. However, there is an 17\% difference between our final result and the currently adopted value~\cite{Sundholm1993JCP,Pekka2018MP}. When constructing the one-body and two-body correlation potentials to describe the core-core and core-valence correlations, all atomic orbitals (1s, 2s, 2p, 3s, and 3p) are designated as active in our CC calculations. In contrast, it is noteworthy that in the finite-element multiconfiguration Hartree-Fock calculation carried out by Sundholm and Olsen, the 1s and 2s orbitals were in a frozen state~\cite{Sundholm1993JCP}. That is to say, their calculations completely neglected the electron correlation effects associated with the 1s and 2s orbitals. This could be one of the factors contributing to the observed 17\% discrepancy. Notably, the three results of the nuclear quadrupole moment we obtained from the 4s3d \(^1\!D_{2}\), 4s4p \(^3\!P_{1}\), and 4s4p \(^3\!P_{2}\) states show remarkable consistency. Therefore, we are confident that our final result is both reliable and accurate. For other values of the nuclear electric quadrupole moment \(Q\) extracted from the data of neutral Ca~\cite{Grundevik1979PRL,Aydin1982ZPA}, the differences are quite obvious, mainly because the theoretically calculated values they used are not accurate enough and have large uncertainties.

In the last two rows, we also list the values of the nuclear electric quadrupole moment \(Q\) extracted from the HFS parameters of the \(4p_{3/2}\) and \(3d_{5/2}\) states of singly-ionized Ca\(^+\)~\cite{Yu2004PRA,Sahoo2009PRA}. Our result aligns with the recommended result for \(4p_{3/2}\)~\cite{Yu2004PRA}, which can be attributed to the large uncertainty of the measured HFS constant $B$ reported in the Ref.~\cite{Silverans1991ZPD}. Our final result is 7.9\% larger than the one reported by Sahoo~\cite{Sahoo2009PRA}.
We also employ the standard RCCSD method to calculate the electric field gradient \(q\) of the \(3d_{5/2}\) state of \(^{43}\)Ca\(^+\), and our result is consistent with the value reported by Sahoo~\cite{Sahoo2009PRA}. However, it is worth noting that the \(nd_{5/2}\) metastable state of the singly-ionized alkaline-earth ion (Ca\(^+\)(n=3), Sr\(^+\)(n=4), and Ba\(^+\)(n=5)) system is very sensitive to electron correlation effects~\cite{Safronova2011PRA,Safronova2010PRA,Porsev2021PRA}. Accurately calculating the hyperfine interaction parameters of the \(nd_{5/2}\) metastable state may require a comprehensive consideration of the contributions from triple excitations, and even quadruple excitations, beyond the CCSD level. As far as we are aware, only one research group has measured the hyperfine splitting of the \(3d_{5/2}\) state~\cite{Benhelm2007PRA}. Notably, the hyperfine splitting of the \(4d_{5/2}\) state of the homologous ion \(^{87}\)Sr\(^+\) was measured using a similar approach, and the HFS constant \(A\) and \(B\) were extracted~\cite{Barwood2003pra}. Sahoo used the RCC method to determine the electric quadrupole moment \(Q\) of the \(^{87}\)Sr nucleus~\cite{Sahoo2006pra}. Subsequently, it was found that the \(Q\) value obtained by Sahoo is approximately 7\% lower than the one extracted based on the HFS parameters of the \(5s5p~^3\!P_1\) and \(5s5p~^3\!P_2\) states in neutral \(^{87}\)Sr atom~\cite{Lu2019pra}. Therefore, the reported experimental values and theoretical results of HFS parameters for the \(3d_{5/2}\) state of \(^{43}\)Ca\(^+\) and the \(4d_{5/2}\) state of \(^{87}\)Sr\(^+\) may need further verification.

\begin{table}
	\caption[ ]{The nuclear quadrupole moment $Q$ (in b) of $^{43}$Ca.}\label{Tab5}
	\begin{ruledtabular}
		\begin{tabular}{clllllllllll}
State             & $Q$ & Reference \\
\hline
4s3d $^1\!D_{2}$      &$-0.0480(10)$         &This work\\
4s4p $^3\!P_{1}$      &$-0.0478(8)$          &This work\\
4s4p $^3\!P_{2}$      &$-0.0480(9)$          &This work\\
Final result          &$-0.0479(6)$           &This work\\
4s4p $^3\!P_{2}$      &$-0.065(20)$          &~\cite{Grundevik1979PRL}\\
4s3d $^1\!D_{2}$      &$-0.062(12)$          &~\cite{Aydin1982ZPA}\\
4s4p $^3\!P_{1,2}$    &$-0.049(5)$           &~\cite{Salomonson1984ZPA,Olsson1982ZPA}\\
4s3d $^1\!D_{2}$      &$-0.0408(8)$          &~\cite{Sundholm1993JCP}\\
$4p_{3/2}$ Ca$^{+}$   &$-0.044(9) $          &~\cite{Yu2004PRA}\\
$3d_{5/2}$ Ca$^{+}$   &$-0.0444(6)$          &~\cite{Sahoo2009PRA}\\
\end{tabular}
\end{ruledtabular}
\end{table}

In conclusion, the nuclear quadrupole moment \(Q = -0.479(6)\)~b, which is extracted from the 4s3d \(^1\!D_{2}\), 4s4p \(^3\!P_{1}\), and 4s4p \(^3\!P_{2}\) states of the \(^{43}\)Ca atom, is the most reliable result to date. We recommend using this result as the new reference value for \(^{43}\)Ca.
\subsection{Various correlation corrections}

Table~\ref{Tab6} presents the contributions of various correlation corrections to HFS constants \(A\) and \(B\) of the \(4s3d\ ^1\!D_{2}\), \(4s4p\ ^3\!P_{1}\), and \(4s4p\ ^3\!P_{2}\) states in the \(^{43}\text{Ca}\) atom. These correlation corrections include the random-phase approximation (RPA) correction, core Brueckner correction, structural radiation correction, normalization correction, and two-particle interaction correction. The calculation results are obtained within the framework of the RCI + LCCSD method

In Table~\ref{Tab6}, "DF" indicates that the single-electron reduced matrix element $o^{(k)}_{vw}$ in Eq.(~\ref{one}) is obtained using the DF wave function. "RPA" stands for the Random Phase Approximation correction, "HO" denotes the cumulative contribution from core Brueckner, structural radiation, and normalization corrections. "TP" corresponds to the contribution brought about by the two-particle interaction.

It can be clearly seen from Table~\ref{Tab6} that the RPA, HO, and TP corrections all play  crucial roles. For the HFS constants \(A\), the contribution of the RPA correction is the most prominent. Notably, the signs of the contributions of the RPA correction and the HO correction are opposite, which leads to a cancellation effect between them. Specifically, for the \(4s3d\ ^1\!D_{2}\) state, the sign of the RPA correction is opposite to those of the other two contributions. For the other two states, the RPA and HO corrections still have opposite signs, while the TP correction has the same sign as the RPA correction. Among these contributions, the RPA correction has the most significant influence, followed by the HO correction, and then the TP correction.
\begin{table}
	\caption[ ]{Contributions of various correlation effects to HFS constants $A$ and $B$ for 4s3d $^1\!D_{2}$, 4s4p $^3\!P_{1}$, and 4s4p $^3\!P_{2}$ states in $^{43}$Ca atom in MHz.}\label{Tab6}
	\begin{ruledtabular}
		\begin{tabular}{crrrrrrrrrrrr}
State	&	DF	&	RPA	&	HO	&	TP	&	Total	\\
\hline
	        &\multicolumn{5}{c}{HFS constants $A$}	 	\\
4s3d $^1\!D_{2}$	&	18.63 	&  $-5.23$ 	&	 2.75 	&	1.29 	&	17.44 	\\
4s4p $^3\!P_{1}$	&	165.28 	&	37.54 	&  $-8.00$ 	&	3.34 	&	198.17 	\\
4s4p $^3\!P_{2}$	&	143.97 	&	33.17 	&  $-7.27$ 	&	1.83 	&	171.7 	\\
	&\multicolumn{5}{c}{HFS constants $B$}	 	\\
4s3d $^1\!D_{2}$	& $-4.057$ 	&	$-0.224$ 	&  $0.060$ 	&	$-0.418$ 	&	$-4.639$ 	\\
4s4p $^3\!P_{1}$	& $1.904$ 	& $0.823$ 	&	 $-0.197$ 	& 0.152 	& 2.682 	\\
4s4p $^3\!P_{2}$	& $-3.736$ 	& $-1.627$ 	&  0.393 	&	$-0.303$ 	&	$-5.273$ 	\\
\end{tabular}
\end{ruledtabular}
\end{table}

For HFS constants \(B\), the signs of the RPA correction and the HO correction are opposite, resulting in mutual cancellation, and the sign of the TP correction is the same as that of the RPA correction. In the \(4s3d\ ^1\!D_{2}\) state, the contribution of the TP correction is quite considerable. Specifically, for the HFS constant \(A\), the TP correction accounts for 7.5\% of the total value; for the HFS constant \(B\), the TP correction accounts for 9.0\% of the total value. For the \(4s4p\ ^3\!P_{1}\) and \(4s4p\ ^3\!P_{2}\) states, although the proportion of the TP correction is not as large as that in the \(4s3d\ ^1\!D_{2}\) state, it is still of great significance, especially for the HFS constants \(B\). In addition, we also find that there is a cancellation phenomenon between the HO correction and the TP correction for the HFS constants \(B\) of the \(4s4p\ ^3\!P_{1}\) and \(4s4p\ ^3\!P_{2}\) states. In previous many-body perturbation calculations, most RCI+MBPT calculations only considered the RPA correction, yet satisfactory results could still be obtained. This is most likely because there is a cancellation effect between the HO correction and the TP correction for these states. In conclusion, to accurately calculate the properties of hyperfine interactions, these corrections should be considered simultaneously.

\section{Conclusion}\label{conclusions}
The primary aim of the present work is to resolve the disparity between the nuclear electric quadrupole moments previously derived from the hyperfine-structure parameters of the neutral \(^{43}\text{Ca}\) atom and the singly-ionized \(^{43}\text{Ca}^+\) ion. To attain this goal, we have developed a code based on a hybrid approach that integrates the advantages of the configuration interaction method and the coupled-cluster method. This hybrid approach can simultaneously account for core-core, core-valence, and valence-valence correlations. Specifically, core-core and core-valence correlations are established through the coupled-cluster with single and double approximations calculation, while valence-valence correlation is considered via the configuration interaction calculation. During the calculation of the transition matrix elements, we comprehensively incorporate effects including the random-phase approximation correction, core Brueckner correction, structural radiation correction, and normalization correction to all orders. Moreover, we take the two-body interaction into account up to the second-order level.

The energies of the low-lying states and the magnetic dipole hyperfine-structure constants of the \(4s3d\ ^1\!D_{2}\), \(4s4p\ ^3\!P_{1}\), and \(4s4p\ ^3\!P_{2}\) states in the neutral \(^{43}\text{Ca}\) atom are calculated. Four different methods were employed to construct the core-core and core-valence correlation potentials, namely MBPT(2), LCCSD, CCSD, and CCSDs. In terms of energy properties, we found that the RCI+MBPT(2) method exhibits the most substantial discrepancy when compared with the experimental values. The RCI+CCSD method outperforms the RCI+LCCSD method in terms of accuracy, achieving at least a two-fold improvement. Regarding the magnetic dipole hyperfine-structure constant, it can be noted that the results obtained by the RCI+MBPT method deviate the most from the experimental values. The results calculated by the RCI+LCCSD, RCI+CCSD, and RCI+CCSDs methods are very close to one another. The result of the RCI+LCCSD method is the closest to the experimental result for the \(^{43}\text{Ca}\) atom, with the maximum difference not exceeding 1.5\%. These calculations confirm that the hybrid method combining the CI method and the CC method can effectively account for the majority of electron correlation effects and provide relatively accurate results.

Then, we applied the same methods to calculate the electric field gradients of the \(4s3d\ ^1\!D_{2}\), \(4s4p\ ^3\!P_{1}\), and \(4s4p\ ^3\!P_{2}\) states in the \(^{43}\text{Ca}\) atom. By combining the measured values of the electric quadrupole hyperfine-structure constants of these three states, we determined the electric quadrupole moment, \(-0.0479(6)\)~b, of the \(^{43}\text{Ca}\) nucleus. This value is 17\% larger than the currently adopted value~\cite{Sundholm1993JCP,Pekka2018MP}. It is also 7.9\% larger than the electric quadrupole moment extracted from \(^{43}\text{Ca}^+\)~\cite{Sahoo2009PRA}. The three electric quadrupole moments obtained from these three states are all consistent with each other. Therefore, we believe that our electric quadrupole moment $Q$ are more reliable than those previously reported~\cite{Grundevik1979PRL, Arnold1981HI,Aydin1982ZPA,Olsson1982ZPA,Sundholm1993JCP,Sahoo2009PRA}. We suggest adopting the current \(Q=-0.0479(6)\)~b presented herein as a reference for \(^{43}\text{Ca}\).
Additionally, the previously reported experimental and theoretical values of HFS parameters for the \(3d_{5/2}\) state of \(^{43}\)Ca\(^+\) may need further verification.

We also analyzed the contributions of the random-phase approximation correction, core Brueckner correction, structural radiation correction, normalization correction, and two-particle interaction corrections to the HFS constants \(A\) and \(B\) of the \(4s3d\ ^1\!D_{2}\), \(4s4p\ ^3\!P_{1}\), and \(4s4p\ ^3\!P_{2}\) states in the \(^{43}\text{Ca}\) atom. We observed that these corrections are all significant, and there are phenomena of opposite signs and cancellation among them. Therefore, high-precision calculation of hyperfine interaction parameters necessitates the simultaneous consideration of these corrections. This analysis is of great importance as it offers guidance for the subsequent application of this method to calculate HFS parameters of other atomic systems.

\begin{acknowledgments}
We are grateful to T.-Y. Shi, Y.-H. Zhang, and Y.-J. Cheng for reading our manuscript.The work was supported by the National Natural Science Foundation of China under Grant No.12174268.
\end{acknowledgments}
%

\end{document}